%% file: MIMO_SC_LE_Final.tex
\documentclass[10pt,twocolumn]{IEEEtran}
\usepackage[dvips]{graphics}
\usepackage{times}
\usepackage{indentfirst}
\usepackage{amsmath}
\usepackage{epsfig}
\usepackage{graphicx}
\usepackage{float}
\usepackage{amsfonts}
\usepackage{amssymb}
\usepackage{multirow}

\begin{document}

\input newcommand.tex

\title{Optimal Tx-BF for MIMO SC-FDE Systems
\author{Peiran Wu, Robert Schober, and Vijay K. Bhargava}
\thanks{The authors are with the Department of Electrical and Computer
Engineering, University of British
Columbia, Vancouver, BC Canada, V6T,
1Z4, email: \{peiranw, rschober,
vijayb\}@ece.ubc.ca. } } \maketitle

\begin{abstract}
Transmit beamforming (Tx-BF) for
multiple-input multiple-output (MIMO)
channels is an effective means to
improve system performance. In
frequency-selective channels, Tx-BF can
be implemented in combination with
single-carrier frequency-domain
equalization (SC-FDE) to combat
inter-symbol interference. In this
paper, we consider the optimal design
of the Tx-BF matrix for a MIMO SC-FDE
system employing a linear minimum mean square error (MSE)
receiver. We formulate the Tx-BF optimization problem as
the minimization of a general function
of the stream MSEs, subject to a
transmit power constraint. The optimal
structure of the Tx-BF matrix is
obtained in closed form and an
efficient algorithm is proposed for
computing the optimal power allocation.
Our simulation results validate the
excellent performance of the proposed
scheme in terms of uncoded bit-error
rate and achievable bit rate.
\end{abstract}


\section{Introduction}

Multiple-input multiple-output (MIMO)
systems are a promising technology to
improve the spectrum efficiency and/or
error performance of wireless networks.
In order to fully exploit the benefits
of the multiple antennas and the
available channel state information at
the transmitter (CSIT), appropriate
MIMO transmit beamforming (Tx-BF)
schemes are required. Optimal MIMO
Tx-BF designs for flat fading channels
and frequency selective fading channels
in combination with orthogonal
frequency-division multiplexing (OFDM)
have been well studied in the
literature, cf.
\cite{BF1}-\cite{Jiang}.
 {However, the design methodology
used in \cite{BF1}-\cite{Jiang} is not directly
applicable to MIMO systems employing
single-carrier frequency-domain
equalization (SC-FDE) \cite{FD-DFE},
where a block circular matrix structure
is imposed on the equalization matrix
to enable efficient frequency domain
implementation. In particular, for MIMO SC-FDE systems, the system
performance metrics depend on the
mean square errors
(MSEs) of the spatial data streams in the time domain, instead of
the subcarrier MSEs in the frequency
domain as is the case for OFDM systems.} Tx-BF design for
SC-FDE systems has been investigated in
several works. For example, adopting
the arithmetic MSE
(AMSE) as the performance metric,
\cite{SC2} proposed optimal Tx-BF for a
MIMO SC-FDE system with both linear and
decision feedback equalization.
However, obtaining the
optimal Tx-BF matrix design directly
minimizing the bit-error rate (BER) or maximizing the
achievable bit rate (ABR) is much more challenging since,
unlike the AMSE, these performance
metrics are nonlinear functions of the
data stream MSEs. In \cite{SC3}, the
authors provide a first attempt to
minimize the BER of a multiple-input
single-output (MISO) SC-FDE system.
However, since only one data stream is
transmitted, minimizing the BER is
equivalent to minimizing the AMSE for
MISO systems. To the best of the
author's knowledge, Tx-BF design for
MIMO SC-FDE systems with general
non-AMSE based objective functions has
not been studied in the literature yet.

In this paper, we propose an optimal
Tx-BF design for MIMO SC-FDE systems.
After deriving the optimal linear
minimum MSE receiver and the associated
stream MSEs, we formulate the Tx-BF
optimization problem as the
minimization of general Schur-convex
and Schur-concave functions of the MSEs
under a transmit power constraint. To
solve the optimization problem, we
first obtain the optimal structure of
the Tx-BF matrix based on majorization
theory. This allows us to transform the
original complex matrix problem into a
real scalar power optimization problem.  {Similar to the case of MIMO OFDM,
the proof of the optimal structure of the Tx-BF matrix for MIMO
SC-FDE is based on the Schur-convexity/concavity of the objective function
\cite{Polo1,Maj}. However, the power allocation problem
for MIMO SC-FDE is quite different from the power allocation
problem for MIMO OFDM.}
 {}

In this paper, $\rm tr(\qA)$, $\qA^{-1}$, $\qA^{T}$, and $\qA^{\dag}$ denote
the trace, inverse, transpose, and conjugate transpose
 of matrix $\qA$, respectively. $\mathbb{C}^{M \times N}$ denotes
 the space of all complex $M \times N$ matrices
 and $\qI_M$ is the $M\times M$ identity matrix.
 $\qn \sim \mathcal{CN}(\mathbf 0 , \sigma_{n}^2\qI_{M})$ indicates
 that $\qn\in \mathbb{C}^{M \times 1}$ is a complex Gaussian
 distributed vector with zero mean and covariance matrix $\sigma_{n}^2\qI_{M}$. $E[\cdot]$
 and $\otimes$ denote statistical expectation and the Kronecker product,
  respectively. $\mbox{blkcirc}([\qA_1^T,\qA_2^T,...,\qA_M^T]^T)$ and
   $\mbox{blkdiag}([\qA_1^T,\qA_2^T,...,\qA_M^T]^T)$ denote a block
   circular matrix and a block diagonal matrix, respectively,
   formed by the block-wise vector
 $[\qA_1^T,\qA_2^T,...,\qA_M^T]^T$. $\mathbb{F} \in \mathbb{C}^{N_c \times N_c}$
 denotes the Fast Fourier Transform (FFT) matrix.

\section{System Model and MSE FDE}

\subsection{System Model}
We consider a MIMO SC-FDE system with
$N_t$ transmit antennas and $N_r$
receive antennas, as shown in Fig.~1.
Let
$\qs_{n}=[s_{n}(1),s_{n}(2),\ldots,s_{n}(M)]^T,
n=0,...,N_c-1,$ denote the symbol
vector at time $n$, where $s_{n}(j)$
denotes the $n$th symbol on the $j$th
spatial stream and $M\leq
\min\{N_t,N_r\}$ is the number of
transmitted data streams. The
$s_{n}(j)$ are independent and
identical distributed with zero mean
and variance $\sigma_{s}^2$. Stacking
all symbol vectors into one vector
leads to $\qs=[\qs_{0}^T,\ldots,
\qs_{N_c-1}^T]^T$. Next, $\qs$ is
transformed into the frequency domain
(FD) and processed by the FD BF matrix
$\qP_f={\rm blkdiag}\{[\qP_{0}^T,...,
\qP_{N_c-1}^T]^T\}$\footnote{ {$\qP_f$
is restricted to be block-diagonal to
enable efficient FD implementation
 of Tx-BF and FDE.}},
with $\qP_k\in \mathbb{C}^{N_t\times
M}$ being the BF matrix at frequency
$k$. Subsequently, the signal is
converted back to the time domain (TD).
Then, the signal is prepended by a
Cyclic Prefix (CP), which includes the
last $N_t K$ data symbols ($K \geq L$
and $L$ is the maximum channel impulse
response (CIR) length), and sent over
the time dispersive channel.

%\begin{figure*}[htp]
%    \centering
%    \includegraphics[width=5in, height=1.1in]{SC-LE-P2P.eps}
%    \caption{System model for the MIMO SC-FDE system. FFT and IFFT denote the
%    fast Fourier transform and inverse fast Fourier transform, respectively.}
%{\label {}}
%\end{figure*}

%\begin{figure*}[htp]
%    \centering
%    \includegraphics[width=5in, height=1.1in]{SC-LE-P2P.eps}
%    \caption{System model for the MIMO SC-FDE system. FFT and IFFT denote the
%     {$N_c$-point FFT} and inverse FFT, respectively.}
%\end{figure*}

\begin{figure*}[htp]
    \centering
    \includegraphics[width=5in, height=1in]{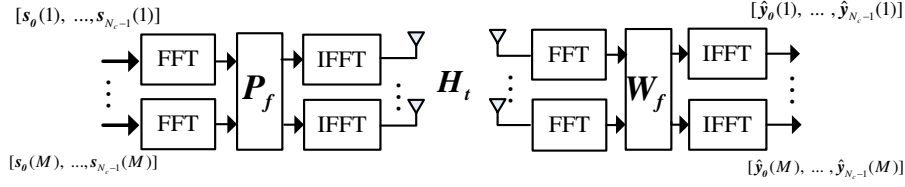}
    \caption{System model for the MIMO SC-FDE system. FFT and IFFT denote the
     {$N_c$-point FFT} and inverse FFT, respectively.}\vspace*{-4mm}
\end{figure*}

The CP converts the linear convolution
of the CIR and the signal vector to a
circular convolution. Hence,
after CP removal, the channel matrix
seen by the receiver is a block
circular matrix $\qH_t={\rm
blkcirc}\{[\qH_{t,0}^T,...,
\qH_{t,L_h-1}^T, \mathbf 0_{N_t\times
N_r(N_c-L_h)}]^T\}$ where $\qH_{t,\ell}
\in \mathbb{C}^{N_r\times N_t}$ denotes
the spatial channel matrix of the
$\ell$th path. Note that the TD channel
matrix can be decomposed as
$\qH_t=\qF_{N_r}^\dag\qH_f\qF_{N_t}$,
where $\qF_X=\mathbb{F}\otimes
\qI_{X}$, and $\qH_f={\rm
blkdiag}\{[\qH_{f,0}^T,...,
\qH_{f,N_c-1}^T]^T\}$ with
$\qH_{f,k}\in \mathbb{C}^{N_r\times
N_t}$ being the FD channel matrix at
frequency $k$. The received signal
after CP removal is then given by \bee
\qy=\qH_t\qF_{N_t}^\dag\qP_f\qF_M\qs+\qn,\eee
where $\qn=[\qn_{1}^T,\ldots,
\qn_{N_c}^{T}]^{T}$ is the noise vector
with $\qn_{n}$ denoting the additive
white Gaussian noise (AWGN) vector at
time $n$. At the receiver, the signal
is converted into the FD and processed
by the FDE matrix $\qW_f={\rm
blkdiag}\{[\qW_{0}^T,...,
\qW_{N_c-1}^T]^T\}$, where $\qW_k\in
\mathbb{C}^{M\times N_r}$ denotes the
FDE matrix at frequency $k$. The
equalized signal in the TD is thus
given by $ \hat
\qy=\qF_{M}^\dag\qW_f\qF_{N_r}\qy$.
Then, the error vector at the equalizer
output is $\qe=\hat\qy-\qs$, and the
corresponding MSE matrix is obtained as
\begin{align}
\label{MSE} & \qE=E[\qe\qe^\dag]=
\qF_M^\dag\Big[\sigma_{s}^2(\qW_f\qH_f\qP_f\qP_f^\dag\qH_f^\dag\qW_f^\dag
\\ &-\qW_f\qH_f\qP_f-\qP_f^\dag\qH_f^\dag\qW_f^\dag
+\qI_{MN_c})+\sigma_{n}^2(\qW_f\qW_f^\dag)\Big]\qF_M.\nnb
\end{align}
%In this work, we assume both the
%transmitter and the receiver has
%perfect CSI to design the optimal Tx-BF
%and FDE matrices. In this case,
%adaptive modulation can be used to
%further improve the performance. We
%shall partially investigate this aspect
%in the simulation results assuming

\subsection{Optimal Minimum MSE FDE}
Based on (2), we can derive the optimal
FDE filter $\qW_f$  {which
minimizes the sum MSE of all spatial streams} for a given
Tx-BF matrix
$\qP_f$\footnote{ {We note that the joint
optimization of $\qW_f$ and $\qP_f$ would lead
to an intractable problem. Hence, as customary
in the literature  \cite{Polo1,Jiang}, we adopt a
suboptimal approach and find first the optimal
minimum MSE FDE filter for a given Tx-BF matrix,
before optimizing the Tx-BF matrix based on a general
objective function that depends on the stream MSEs.}}. By
differentiating ${\rm tr}(\qE)$ with
respect to (w.r.t.) $\qW_f$ and setting
the result to zero, we obtain the
optimal minimum MSE FDE filter and the
corresponding MSE matrix, \bee
\qW_{f}&=&\sigma_{s}^2 \mathbf
\Psi_f^{-1}
\qP_{f}^{\dag}\qH_{f}^{\dag} \quad
\mbox{and} \quad \qE = \sigma_{s}^2
\qF_M^\dag\mathbf \Psi_{f}^{-1}\qF_M,
 \eee respectively, where $ \mathbf
\Psi_f=\frac{\sigma_s^2}{\sigma_n^2}\qP_f^{\dag}\qH_f^{\dag}\qH_f\qP_f+\qI_{MN_c}$.
The $k$th block diagonal matrix entry
of $\qW_f$ is given by
$\qW_{k}=\sigma_{s}^2\mathbf
\Psi_k^{-1}
\qP_{k}^{\dag}\qH_{f,k}^{\dag}$, where
$\mathbf \Psi_k=
\frac{\sigma_s^2}{\sigma_n^2}\qP_k^{\dag}\qH_{f,k}^{\dag}\qH_{f,k}\qP_k+\qI_M$.
Note that since MSE matrix $\qE$ is a
block circular matrix, its block
diagonal entries are all identical,
i.e., $\qE_{k}=\hat \qE,\forall k$. By
exploiting the structure of $\qE$,
$\hat \qE$ can be expressed as
\begin{align} \hat\qE=
\frac{\sigma_{s}^2}{N_c}\sum_{k=0}^{N_c-1}\mathbf
\Psi_{k}^{-1}. \end{align}

\section{Optimal Tx-BF Matrix Design}

Now, we are ready to derive the optimal
$\qP_f$ such that a general function of
the stream MSEs is minimized, under a
power constraint on the Tx-BF matrix.
Mathematically, the optimization
problem is formulated as:
     \bee \label{prob0}
        && \min_{{\rm tr}(\qP_f\qP_f^\dag)\leq P_T}
         \quad
         f(\mbox{diag}[\hat\qE]),\eee
 where $P_T$ is the
 power budget for the transmitter, and $\mbox{diag}[\qM]$ denotes a vector
 containing the diagonal entries of matrix $\qM$. The considered objective
functions $f(\mbox{diag}[\hat \qE])$
can be either Schur-convex or
Schur-concave functions \cite{Polo1}
w.r.t. $\mbox{diag}[\hat\qE]$.

\subsection{Optimal Structure of the Tx-BF Matrix} We
first investigate the optimal structure
of the Tx-BF matrix. We begin by
introducing the singular-value decomposition (SVD) of
the FD channel matrix \bee
\qH_{f,k}=\qU_{H}^{(k)}\mathbf
\Lambda_{H}^{(k)} \qV_{H}^{(k)\dag},
\quad \forall k,  \eee where
$\qU_{H}^{(k)}\in \mathbb{C}^{N_r\times
N_r}$ and $\qV_{H}^{(k)}\in
\mathbb{C}^{N_t\times N_t}$ are the
singular-vector matrices of
$\qH_{f,k}$, and $\mathbf
\Lambda_{H}^{(k)} \in
\mathbb{C}^{N_r\times N_t}$ is the
singular-value matrix of $\qH_{f,k}$
with increasing diagonal elements.
\begin{theo} For the optimization problem in (\ref{prob0}), the following structure of $\qP_k$
is optimal\begin{align}
\label{structure1} \qP_k&=
\bar{\qV}_{H}^{(k)} \mathbf
\Lambda_{P}^{(k)}\qV_{0},\quad \forall
k, \end{align} where
$\bar{\qV}_{H}^{(k)}\in
\mathbb{C}^{N_t\times M}$ contains the
$M$ right-most columns of
${\qV}_{H}^{(k)}$, and $\mathbf
\Lambda_{P}^{(k)}\in\mathbb{C}^{M\times
M}$ is a diagonal matrix with the $m$th
diagonal element denoted by
$\sqrt{P_{km}}$. For Schur-concave
functions, $\qV_{0}=\qI_M$, and for
Schur-convex functions,  $\qV_{0}$ is a
unitary matrix which makes all diagonal
entries of $\hat\qE$ equal \footnote{In
practice, $\qV_{0}$ can be chosen as an
FFT matrix or a Hadamard matrix with
appropriate dimensions.}.
\end{theo}

\begin{proof}
 {Please refer to Appendix A.}
\end{proof}

In the following, we consider some
typical objective functions that are
based on the stream MSEs, and which
have been extensively investigated for
MIMO OFDM systems based on the
subcarrier MSEs \cite{Polo1}.
Specifically, we consider the AMSE,
geometric MSE (GMSE), maximum MSE
(maxMSE), arithmetic
signal-to-interference-plus-noise ratio
(ASINR), geometric SINR (GSINR),
harmonic SINR (HSINR), and arithmetic
bit error rate (ABER) for optimization, i.e.,
\begin{align} \label{obj} f^{\rm [X]}(\mbox{diag}[\hat \qE])= \begin{cases}\begin{array}{c} \sum_{m=1}^{M} \hat\qE_{mm}^{}, \quad \mbox{X=AMSE} \\
                                                    \prod_{m=1}^{M} \hat\qE_{mm}^{},\quad \mbox{X=GMSE}  \\
                                                    \max_{m=1}^{M} \hat\qE_{mm}^{}, \quad \mbox{X=maxMSE} \\
                                                    -\sum_{m=1}^{M} \left(\hat\qE_{mm}^{-1}-1\right), \quad \mbox{X=ASINR} \\
                                                    -\prod_{m=1}^{M} \left(\hat\qE_{mm}^{-1}-1\right), \quad \mbox{X=GSINR} \\
                                                    \sum_{m=1}^{M} \left(\hat\qE_{mm}^{-1}-1\right)^{-1}, \quad \mbox{X=HSINR}\\
                                                    \sum_{m=1}^{M} \alpha Q\left(\sqrt{\beta(\hat\qE_{mm}^{-1}-1)}\right),  \mbox{X=ABER} \\
                                                   \end{array} \end{cases}\nnb \end{align}
where $\hat \qE_{mm}^{}$ is the $m$th
diagonal entry of $\hat\qE$, $Q(\cdot)$ is
the Gaussian $Q$-function, and $\alpha$
and $\beta$ are constellation dependent
coefficients. Here, we assume all
streams adopt the same signal
constellation. Note that the objective functions for
the AMSE, GMSE, ASINR, and GSINR
criteria are Schur-concave functions,
while those for the maxMSE, HSINR, and
ABER criteria are Schur-convex
functions w.r.t. $\mbox{diag}[\hat\qE]$
\cite{Polo1}.

Exploiting the optimal structure of the
Tx-BF matrix in (\ref{structure1}), we
can express $\mathbf {\Psi}_{k}$ as
 $\mathbf {\Psi}_{k}=\qV_0^\dag \mathbf{\hat\Psi}_{k}\qV_0$, with $\mathbf
{\hat\Psi}_{k}=\frac{\sigma_s^2}{\sigma_n^2}\mathbf
\Lambda_{P}^{(k)\dag}\bar{
\mathbf\Lambda}_{H}^{(k)\dag}\bar{
\mathbf\Lambda}_{H}^{(k)}\mathbf\Lambda_{P}^{(k)}
+\qI_M$, where $\bar{
\mathbf\Lambda}_{H}^{(k)}=\mbox{diag}[\sqrt{H_{km}}]$
contains the $M$ largest diagonal
entries of $\mathbf\Lambda_{H}^{(k)}$.
Now, we can write the MSE matrix as
$\hat \qE=
\frac{\sigma_s^2}{N_c}\sum_{k=0}^{N_c-1}\mathbf
{\hat\Psi}_{k}^{-1}$, where the
diagonal entries of $\mathbf
{\hat\Psi}_{k}$ are given by
${\Psi}_{km}=
\frac{\sigma_s^2}{\sigma_n^2}P_{km}
H_{km}+1, \forall m$.

 {Proposition 1:} The
optimization problems in (\ref{prob0})
with Schur-convex objective functions,
i.e., maxMSE, HSINR, and ABER, are
equivalent to AMSE minimization up to a
unitary rotation of the Tx-BF.

\begin{proof} For Schur-convex
functions, the unitary rotation matrix
$\qV_0$ has to make all stream MSEs
identical, e.g., equal to $\hat E$.
Explicitly, we can write $\hat E={\rm
tr}(\hat
\qE)/M=\frac{\sigma_s^2}{MN_c}{\rm
tr}(\sum_{k=1}^{N_c}\mathbf{
\hat{\Psi}}^{-1}_k)=\frac{\sigma_s^2}{MN_c}\sum_{k=1}^{N_c}\sum_{m=1}^{M}{\Psi}_{km}^{-1}$,
which is essentially the objective
function for the AMSE criterion. On the
other hand, as the Schur-convex
objective functions are monotonically increasing w.r.t. $\hat
E$, the corresponding problems are
equivalent to AMSE minimization up to a
unitary rotation (as the AMSE objective
function is Schur-concave).
\end{proof}

 {Remark 1:} For MIMO OFDM
systems, optimization problems
employing different Schur-convex
functions are not equivalent \cite{Polo1},
since in this case, the unitary matrix only
balances the MSEs on each subcarrier,
while the MSEs across the subcarriers
are not identical.

Now, we can restate the objective functions in terms of the new variables
$\qP=\{P_{km},\forall k,m\}$ as $
f^{\rm
[X]}(\qP)=\sum_{m=1}^{M}f_{m}^{\rm
[X]}(\qP)$ with \bee \label{obj} f^{\rm
[X]}_{m} (\qP)= \begin{cases}\begin{array}{c} \frac{\sigma_s^2}{N_c}\sum_{k=0}^{N_c-1}{\Psi}_{km}^{-1}  , \quad \mbox{X=AMSE} \\
                                                    \log_2\Big(\frac{\sigma_s^2}{N_c}\sum_{k=0}^{N_c-1}{\Psi}^{-1}_{km}\Big),\quad \mbox{X=GMSE}  \\
                                                    - \Big(\frac{\sigma_s^2}{N_c}\sum_{k=0}^{N_c-1}{\Psi}^{-1}_{km}
                                                    \Big)^{-1}, \quad \mbox{X=ASINR} \\
                                                    -\log_2 \left(\Big[\frac{\sigma_s^2}{N_c}\sum_{k=0}^{N_c-1}{\Psi}^{-1}_{km}
                                                    \Big]^{-1}-1\right),  \mbox{X=GSINR}
                                                   \end{array} \end{cases} \nnb\eee
 where for GMSE and GSINR, we have taken the logarithm of the original objective functions to facilitate the subsequent optimization. Furthermore, we have omitted
 the objective functions for the maxMSE, HSINR, and ABER criteria since they yield the same power allocation as the AMSE criterion, cf.~Proposition 1.

\subsection{Optimal Power Allocation for the Tx-BF Matrix}
Using Theorem 1, the power constraint
can be expressed
 as $\sum_{k=0}^{N_c-1}\sum_{m=1}^{M} P_{km}
\leq P_T $. The problem is then
reformulated as \bee \label{prob1}
\min_{\sum_{k=0}^{N_c-1}\sum_{m=1}^{M}P_{km}\leq
P_T} && f^{\rm [X]}\left(\qP\right).
\eee  {Proposition 2:}  The
considered objective functions $f^{\rm
[X]}(\qP), {\rm X}=\{\rm AMSE, GMSE,
ASINR, GSINR\}$, are all convex
functions w.r.t. $P_{km}$.

\begin{proof}
Please refer to Appendix B.
\end{proof}
 {The convexity of the problem in
(\ref{prob1}) guarantees the existence
of a global optimum solution for the
power allocation. In addition, since
the power constraint is affine and feasible, the
Slater condition is satisfied
\cite{cvx-book}, implying that strong
duality holds. This allows us to solve
the original primal problem by solving its
dual problem.} To this end, we first
write the Lagrangian of (\ref{prob1})
as $\mathcal{L}= \sum_{m=1}^{M}\left(
f^{\rm [X]}_{m}(\qP) +\lambda
\sum_{k=0}^{N_c-1}P_{km}
\right)-\lambda P_{T}$, where $\lambda$
is the Lagrange multiplier for the
constraint in (\ref{prob1}). The
corresponding dual problem can be
written as \bee
\label{Dual-P}\max_{\lambda\geq
0}~~\min_{\{P_{km}\geq 0\}}~~
\mathcal{L}.\eee  {For a given
$\lambda$, the inner minimization
problem in (\ref{Dual-P}) can be solved
by} applying the Karush-Kuhn-Tucker
conditions \cite{cvx-book}. The optimal
solution for $P_{km}$ is found as \bee
P_{km} =\left( \sqrt{\frac{
\sigma_n^2B_{m}}{\sigma_s^2\lambda
H_{km}}}-\frac{\sigma_n^2}{\sigma_s^2H_{km}}\right)^{+},\label{solu1}
  \eee
  where $B_m$ is a factor that depends on the objective functions,
\begin{align} B_m=
\begin{cases}\begin{array}{c} 1, \quad \mbox{X=AMSE} \\
        (\frac{\sigma_s^2\ln2}{N_c}\sum_{k=0}^{N_c-1}{\Psi}^{-1}_{km})^{-1},\quad \mbox{X=GMSE} \\
        (\frac{\sigma_s^2}{N_c}\sum_{k=0}^{N_c-1}{\Psi}^{-1}_{km})^{-2},\quad \mbox{X=ASINR}\\
        \frac{C_m}{\ln 2}(\frac{\sigma_s^2}{N_c}\sum_{k=0}^{N_c-1}{\Psi}^{-1}_{km})^{-2}, \quad \mbox{X=GSINR}\\
                                                  \end{array} \end{cases} \end{align}
                                                  with $C_m= \left(\Big[\frac{\sigma_s^2}{N_c}\sum_{k=0}^{N_c-1}{\Psi}^{-1}_{km}
                                                    \Big]^{-1}-1\right)^{-1}$.

 {Remark 2: } Recall that for
MIMO OFDM systems \cite{Polo1}, the
Tx-BF optimization based on the GSINR
criterion leads to equal power
allocation, the ASINR criterion leads
to allocating all the transmit power to
the strongest spatial stream among all
the subcarriers, and for Schur-convex
functions, different multilevel
waterfilling solutions are needed. In
contrast, for MIMO SC-FDE systems, the
solutions for all criteria exhibit a
simple single-level waterfilling
structure, cf. (\ref{solu1}).

 {For the outer maximization
problem in (\ref{Dual-P}), we can
obtain the optimal Lagrange multiplier
$\lambda$} by using the iterative
subgradient method
\cite{cvx-book}\begin{align}
\lambda^{[i+1]}=\left[\lambda^{[i]}-
\varepsilon^{[i+1]}\left(\sum_{k=0}^{N_c-1}\sum_{m=1}^{M}P_{km}^{[i]}-P_T\right)\right]^{+},
\end{align}  {where $\varepsilon^{[i]}$ is the step size adopted in the $i$th iteration}, $\lambda^{[i]}$ is the Lagrange multiplier obtained in the $i$th iteration and $P_{km}^{[i]}$ is the solution of (\ref{solu1})
for a given $\lambda^{[i]}$.  {For a nonsummable diminishing step size, i.e., $\lim_{i\rightarrow \infty}\varepsilon^{[i]}=0$, $\sum_{i=1}^{\infty}\varepsilon^{[i]}=\infty$, the subgradient method is
guaranteed to converge to the optimal dual variable $\lambda$ \cite{subG}. The optimal primal variables $P_{km}$ can then be obtained from (\ref{solu1}).}

\section{Simulation Results}

In this section, we evaluate the
performance of the proposed Tx-BF
schemes for MIMO SC-FDE using
simulations. Each data block contains
$N_c=64$ symbols. The channel vectors
are modeled as uncorrelated Rayleigh
block fading channels with power delay
profile \cite{Ch-book} $
p[n]=\frac{1}{\sigma_t}\sum_{l=0}^{L-1}
e^{-n/\sigma_t}\delta[n-l]$,  where
$\sigma_t=2$, which corresponds to
moderate frequency-selective fading.
For convenience, we assume $L$ and $K$
are both equal to 16. The number of
spatial data streams and the number of
transmit and receive antennas are all
set to be two, i.e., $M=N_t=N_r=2$.
The signal to noise ratio is defined as
$SNR=\frac{\sigma_s^2 P_T}{M
N_c \sigma_n^2}$. All
simulations are averaged over at least
100,000 independent channel
realizations and data blocks.

In Fig.~2, we show the uncoded BER of a
MIMO SC-FDE system with quaternary phase shift keying (QPSK) transmission using different
optimization criteria.  {For comparison, the BER of a MIMO SC-FDE system with equal power allocation (EPA)} and
the BERs of optimized MIMO OFDM systems are also shown. As can be seen, MIMO
SC-FDE achieves a much lower BER than
its OFDM counterpart since SC-FDE can
exploit the frequency-diversity of the
channel. Similar to MIMO OFDM, SC-FDE
systems optimized for Schur-convex
functions perform better than those
optimized for Schur-concave functions.
For SC-FDE systems, all Schur-convex
objective functions lead to the same
BER, which is much lower than that for
the Schur-concave AMSE criterion
although they use identical power
allocations.

In Fig.~3, we show the average {achievable bit rate (ABR)} of MIMO
SC-FDE employing the proposed Tx-BF
scheme. The corresponding ABR of
an optimized MIMO OFDM system are also
shown for reference.
For both systems, we have assumed perfect channel loading with
continuous constellation size and optimal channel coding \cite{FD-DFE}.
From the figure, we observe that the GMSE criterion achieves the highest
ABR and GSINR only suffers from a
small ABR loss in the low SNR
regime compared to GMSE. In fact, it is
straightforward to show that GMSE
minimization is equivalent to ABR
maximization. The SNR gap between
the best SC-FDE and OFDM ABR
curves is around 1 dB. Different from
OFDM, for SC-FDE, the systems optimized
for Schur-convex functions achieve the
same ABRs as the system optimized
for the AMSE criterion, which implies
that the unitary rotation matrix
$\qV_0$ does not affect the ABR
performance. {Furthermore, for the
maxMSE and ABER criteria, SC-FDE
systems achieve much higher ABRs
than OFDM systems, as the latter
allocate most of the available power to weaker
subcarriers to balance the MSE/BER, and thus compromise the system ABR.}

\begin{figure}[htp]
    \centering
    \includegraphics[width=3.5in]{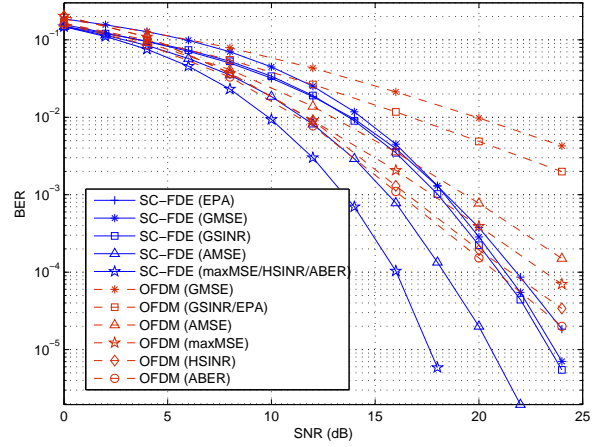}
        \vspace*{-5mm}
    \caption{BER comparison of the optimized MIMO SC-FDE and MIMO OFDM systems.}
{\label {}}
\vspace*{-2mm}
\end{figure}

\begin{figure}[htp]
    \centering
    \includegraphics[width=3.5in]{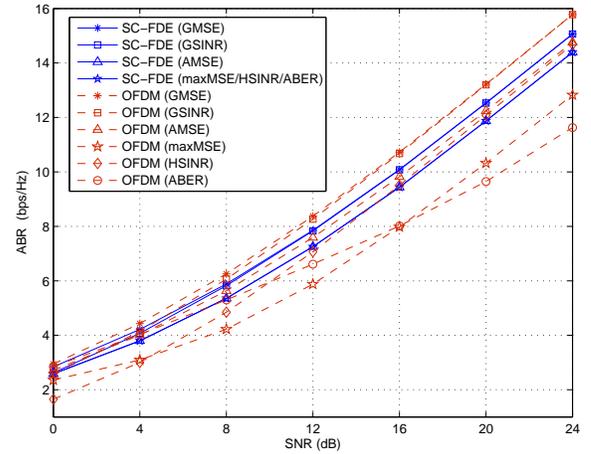}
    \vspace*{-5mm}
    \caption{ABR comparison of the optimized MIMO SC-FDE and MIMO OFDM systems.}
{\label {}}
\vspace*{-2mm}
\end{figure}

\section{Conclusion}
In this paper, we addressed the problem
of Tx-BF matrix design for MIMO SC-FDE
systems. The optimal minimum MSE FDE
filter was derived first, along with
the stream MSEs at the equalizer
output. Then, we optimized the Tx-BF
matrix for minimization of a general
function of the stream MSEs. We found the
optimal structure of the Tx-BF matrix
in closed form and proposed an
efficient algorithm to solve the
remaining power allocation problem. Our
results show that AMSE based Tx-BF
optimization \cite{SC2,SC3} is neither
optimal for minimizing the uncoded BER
nor for maximizing the achievable bit
rate of MIMO SC-FDE systems.

\section*{Appendix A}
For Schur-concave functions, from
majorization theory [7, 9.B.1], we have
$f({\rm diag}(\hat\qE))\geq f({\rm
eig}(\hat\qE))$, where ${\rm
eig}(\hat\qE)$ is a vector containing
the eigenvalues of $\hat\qE$ sorted in
decreasing order and equality is
achieved if $\hat\qE$ is a diagonal
matrix. From (4), a sufficient
condition to achieve this objective
function lower bound is to make
$\mathbf \Psi_k^{-1}, \forall k,$
diagonal, i.e.,
$\qP_k^\dag\qH_{f,k}^\dag\qH_{f,k}\qP_k=\qD_k
\in \mathbb{C}^{M \times M}$ is a
diagonal matrix with increasing
diagonal entries. For subsequent use,
we write
$\qH_{f,k}\qP_k=\qQ_k\qD_k^{1/2}$,
where $\qQ_k\in \mathbb{C}^{N_t \times
M}$ is a unitary matrix, and
$\qD_k^{1/2}$ is the square root of
$\qD_k$. Now, we derive the optimal
$\qP_k$ that minimizes the transmit
power. {For notational simplicity, we
only} consider the case when ${\rm
rank}(\qH_{f,k})=N_t=N_r, \forall k$.
Using the SVD of $\qH_{f,k}$, we can
rewrite $\qP_k$ as
$\qP_k=\qV_H^{(k)}\mathbf
\Lambda_H^{(k)-1}\qU_H^{(k)\dag}\qQ_k\qD_k^{1/2}$.
The power consumption at frequency $k$
is thus given by ${\rm
tr}(\qP_k\qP_k^\dag)={\rm tr}(\mathbf
\Lambda_H^{(k)-2}\qT_k\qD_k\qT_k^\dag)\geq
{\rm tr}(\mathbf
{\bar\Lambda}_H^{(k)-2}\qD_k)$, where
$\mathbf {\bar\Lambda}_H^{(k)}$ is a
diagonal matrix containing the $M$
largest singular values of $\mathbf
{\Lambda}_H^{(k)}$,
$\qT_k=\qU_H^{(k)\dag}\qQ_k$, and the
inequality is due to [7, 9.H.1]. Since
equality is achieved if $\qT_k=[
\mathbf 0_{M\times (N_T-M)}\quad
\qI_{M}]^T$, we obtain
$\qQ_k=\bar\qU_H^{(k)}$, where
$\bar\qU_H^{(k)}$ contains the $M$
right-most columns of $\qU_H^{(k)}$.
Plugging $\qQ_k=\bar\qU_H^{(k)}$ into
the expression for $\qP_k$, we obtain
the optimal $\qP_k$  as
$\qP_k=\bar\qV_H^{(k)}\mathbf
\Lambda_P^{(k)}$ where
$\bar\qV_H^{(k)}$ contains the $M$
right-most columns of $\qV_H^{(k)}$, and
$\mathbf \Lambda_P^{(k)}=\mathbf
{\bar\Lambda}_H^{(k)-1}\mathbf
\qD_k^{1/2}$.

For Schur-convex objective functions,
from [7, p.7], we have $f({\rm
diag}(\hat\qE))\geq f({\rm
tr}(\hat\qE)\mathbf 1/M)$, where
$\mathbf 1$ is the all-one vector, and
equality is achieved if $\hat\qE$ has
identical diagonal entries equal to
${\rm tr}(\hat\qE)/M$. Denote the
eigenvalue decomposition of $\hat
\qE(\{\qP_k\})$ for a feasible $\qP_k$
as $\hat
\qE(\{\qP_k\})=\qU_E^\dag\mathbf
\Lambda_E\qU_E$. Now consider another
feasible Tx-BF matrix $\tilde
\qP_k=\qP_k\qU_E\qV_0$, where $\qV_0$
is a unitary matrix. Then, we have
$\hat
\qE(\{\tilde\qP_k\})=\qV_0^\dag\mathbf
\Lambda_E\qV_0$ and according to [7,
9.B.2], we can always find a $\qV_0$
such that $\hat \qE(\{\tilde\qP_k\})$
has identical diagonal entries which
equal ${\rm
tr}(\hat\qE(\{\tilde\qP_k\})/M)$. Since
$f({\rm diag}[{\rm tr}(\hat\qE)\mathbf
1/M])$ is monotonic with respect to its
argument, the remaining task is to
minimize ${\rm tr}(\hat\qE)$, which is
a Schur-concave function. Since for
Schur-concave function, the optimal
$\qP_k$ diagonalizes $\hat \qE$, we
have $\qU_E=\qI_M$. The optimal Tx-BF
matrix for Schur-convex objective
functions can then be obtained as
$\tilde\qP_k=\qP_k\qV_0=\bar\qV_H^{(k)}\mathbf
\Lambda_P^{(k)}\qV_0$. This completes
the proof.

\section*{Appendix B}
For the AMSE criterion, the second
order derivatives of the objective
function are given by $\frac{\partial^2
f_m^{\rm [AMSE]}(\qP)}{\partial
 P_{km}^2}=2{\Psi}^{-3}_{km}H_{km}^2\geq
 0$ and $\frac{\partial^2 f_m^{\rm[AMSE]}(\qP)}{\partial
 P_{km}P_{jm}}=0, \forall j\neq k$.
 Therefore, the Hessian matrix is a
 diagonal matrix with non-negative
 diagonal entries, which implies that
 $f_m^{\rm
[AMSE]}(\qP)$ is a convex function
w.r.t. $P_{km}, \forall k,m$. For the
GMSE criterion, we rewrite the
objective function as $f_m^{\rm
[GMSE]}(\qP)=\log_2\Big(\frac{\sigma_s^2}{N_c}\sum_{k=0}^{N_c-1}\exp
(-\log{\Psi}_{km})\Big)$, since
$-\log{\Psi}_{km}$ is a convex
function, and
$\log_2\Big(\frac{\sigma_s^2}{N_c}\sum_{k=0}^{N_c-1}\exp
y\Big)$ is a convex increasing function
w.r.t. $y$ \cite{cvx-book}, the
composition of the two, i.e., $
f_m^{\rm [GMSE]}(\qP)$, is also a
convex function w.r.t. $P_{km}, \forall
k,m$. For the SINR related criteria, it
can be shown that the Hessian matrix of
the stream SINR, i.e.,
$(\frac{1}{N_c}\sum_{k=0}^{N_c-1}{\Psi}_{km}^{-1})^{-1}-1$,
is negative semidefinite. Hence, the
stream SINRs are concave functions
w.r.t. $P_{km}, \forall k,m$. Based on
this concavity, it is straightforward
to show that the objective functions for
ASINR and GSINR maximization are
convex functions w.r.t. $P_{km},
\forall k,m$.

{}
\end{document}

%% file: newcommand.tex
\newtheorem{pro}{Proposition}
\newtheorem{prop}{Property}
\newtheorem{theo}{Theorem}
\newtheorem{lem}{Lemma}
\newtheorem{rem}{Remark}

 \newcommand{\bee}{\begin{eqnarray}}
 \newcommand{\eee}{\end{eqnarray}}

 \newcommand{\be}{\begin{align}}
  \newcommand{\ee}{\end{align}}

 \newcommand{\nnb}{\nonumber}
 \newcommand{\qP}{ {\mathbf P}}
 \newcommand{\cf}{ {$ \,_1F_1 $}}
 \newcommand{\gf}{ {$ \,_2F_1 $}}
 \newcommand{\hf}{ {$\frac{1}{2}$}}

 \newcommand{\Ea}{ {$\frac{\Omega_1  t^2}{4}$}}
 \newcommand{\Eb}{ {$\frac{\Omega_2  t^2}{4}$}}
 \newcommand{\Ec}{ {$\frac{\Omega_3  t^2}{4}$}}
 \newcommand{\Ek}{ {$\frac{\Omega_k  t^2}{4}$}}

 \newcommand{\ppa}{ {$\sqrt{\frac{\pi \Omega_1}{4}}$}}
 \newcommand{\ppb}{ {$\sqrt{\frac{\pi \Omega_2}{4}}$}}
 \newcommand{\ppc}{ {$\sqrt{\frac{\pi \Omega_3}{4}}$}}
 \newcommand{\qqk}{ {$\sqrt{\frac{\pi \Omega_k}{4}}$}}

\newcommand{\qq}{{\mathbf q}}
\newcommand{\qT}{{\mathbf T}}

 \newcommand{\qsum}{ {$\Omega_1 + \Omega_2 +\Omega_3$}}
 \newcommand{\qii}{ {$\int_{-\infty}^{\infty}$}}
 \newcommand{\qa}{ {\mathbf a}}
 \newcommand{\qb}{ {\mathbf b}}
 \newcommand{\qc}{ {\mathbf c}}
 \newcommand{\qe}{ {\mathbf e}}

 \newcommand{\qg}{ {\mathbf g}}
 \newcommand{\qh}{ {\mathbf h}}
 \newcommand{\qm}{ {\mathbf m}}
 \newcommand{\qn}{ {\mathbf n}}

 \newcommand{\qp}{ {\mathbf p}}
 \newcommand{\qr}{ {\mathbf r}}
 \newcommand{\qs}{ {\mathbf s}}
 \newcommand{\qt}{ {\mathbf t}}
 \newcommand{\qu}{ {\mathbf u}}

 \newcommand{\qv}{ {\mathbf v}}
 \newcommand{\qw}{ {\mathbf w}}
 \newcommand{\qx}{ {\mathbf x}}
 \newcommand{\qy}{ {\mathbf y}}
 \newcommand{\qz}{ {\mathbf z}}

 \newcommand{\qA}{ {\mathbf A}}
 \newcommand{\qB}{ {\mathbf B}}
 \newcommand{\qC}{ {\mathbf C}}
 \newcommand{\qD}{ {\mathbf D}}
 \newcommand{\qE}{ {\mathbf E}}
 \newcommand{\qF}{ {\mathbf F}}
 \newcommand{\qG}{ {\mathbf G}}
 \newcommand{\qH}{ {\mathbf H}}
 \newcommand{\qI}{ {\mathbf I}}
 \newcommand{\qL}{ {\mathbf L}}
 \newcommand{\qM}{ {\mathbf M}}
 \newcommand{\qQ}{ {\mathbf Q}}

 \newcommand{\qR}{ {\mathbf R}}
 \newcommand{\qS}{ {\mathbf S}}
 \newcommand{\qX}{ {\mathbf X}}
 \newcommand{\qY}{ {\mathbf Y}}
 \newcommand{\qZ}{ {\mathbf Z}}
 \newcommand{\qU}{ {\mathbf U}}
 \newcommand{\qV}{ {\mathbf V}}
 \newcommand{\qW}{ {\mathbf W}}
 \newcommand{\qK}{ {\mathbf K}}
 \newcommand{\zR}{ {\cal R}}
 \newcommand{\qLambda}{ {\mathbf \Lambda}}
 \newcommand{\qrho}{ {$\boldsymbol \rho$}}
 \newcommand{\qzero}{ {\mathbf 0}}

\newcommand{\hs}[2]{ {${\hat{#1}}{_{#2}}$}}
\newcommand{\bs}[2]{ {${\mathbf {#1}}{_{#2}}$}}
\newcommand{\hbs}[2]{ {${\hat{\mathbf {#1}}}{_{#2}}$}}
\newcommand{\hb}[1]{ {${\hat{\mathbf #1}}$}}
\renewcommand{\thesection}{\Roman{section}}